# Design Priorities in Digital Gateways: A Comparative Study of Authentication and Usability in Academic Library Alliances

**Authors:** Rui Shang, Bingjie Huang

Office of Information Technology (Library), Westlake University, Hangzhou, China

Email: shangrui@westlake.edu.cn, huangbingjie@westlake.edu.cn

## Abstract

**Purpose:** This study examines the design and functionality of university library login pages across academic alliances (IVY Plus, BTAA, JULAC, JVU) to identify how these interfaces align with institutional priorities and user needs. It explores consensus features, design variations, and emerging trends in authentication, usability, and security.

**Methodology:** A multi-method approach was employed: screenshots and HTML files from 46 institutions were analyzed through categorization, statistical analysis, and comparative evaluation. Features were grouped into authentication mechanisms, usability, security/compliance, and library-specific elements.

**Findings:** Core functionalities (e.g., ID/password, privacy policies) were consistent across alliances. Divergences emerged in feature emphasis: mature alliances (e.g., BTAA) prioritized resource accessibility with streamlined interfaces, while emerging consortia (e.g., JVU) emphasized cybersecurity (IP restrictions, third-party integrations). Usability features, particularly multilingual support, drove cross-alliance differences. The results highlighted regional and institutional influences, with older alliances favoring simplicity and newer ones adopting security-centric designs.

**Originality/Value:** This is the first systematic comparison of login page designs across academic alliances, offering insights into how regional, technological, and institutional factors shape digital resource access. Findings inform best practices for balancing security, usability, and accessibility in library interfaces.

# Introduction

University libraries are evolving to meet the dynamic needs of researchers and students in the digital age. In recent years, resource sharing and open access initiatives, often facilitated by library associations, have gained popularity as they are crucial for expanding access to information while limiting growing financial burdens (Cohen et al., 2024). These initiatives democratize information access and enable institutions to collaboratively support scholarly communication. For example, library consortia such as JULAC and BTAA have implemented shared resource systems to enhance accessibility and reduce redundancy in collection development (Pionke & Schroeder, 2020). As the increasing usage of online discovery services, this evolution in collaboration is bound to require a focus on authentication, and UI design. Firstly, effective access management ensure secure and seamless interactions with library resources. Federated identity management (FIM) systems like Shibboleth or OpenAthens have been widely adopted to streamline user authentication while maintaining privacy. These systems simplify access pathways and reduce user frustration caused by inconsistent login experiences across platforms (Felts, 2023). Additionally, Research shows that well-designed interfaces significantly improve behavioral engagement among diverse users (Raut & Singh, 2024). For instance, libraries that incorporate universal design principles—such as multi-language support, responsive layouts, and accessible navigation—are better equipped to serve diverse user populations in libraries (Redkina, 2024 ) Being the inevitable gate to access collection resources, the login page is a combination of authentification technology and interface design, which should reveal the rationale and values behind cooperation ( Fei& Esteban, 2024).

Following the implementation of resource-sharing initiatives, library login page design has emerged as a critical component of the overall user experience in accessing academic resources. While there is broad consensus within academic libraries regarding authentication technologies—with OpenAthens being widely adopted as a standardized framework (Romano & Huynh, 2021)—research on the functional design of login interfaces remains underexplored. Authentication mechanisms, though integral to login pages, have achieved relative stability and uniformity. Notably, SAML (Security Assertion Markup Language) has become a cornerstone in academic library authentication systems due to its support for Single Sign-On (SSO), enabling seamless cross-domain authentication—a necessity for libraries interacting with diverse external resources such as academic databases and journal platforms (Ruenz, 2022).

In contrast, the visual and functional design of library login pages exhibits significant variability, with minimal standardization. Existing scholarship has predominantly focused on holistic website design and functionality, often neglecting login page specifics (Madhusudhan & Nagabhushanam, 2012). While prior studies (e.g., Abifarin et al., 2019; Tella & Oladapo, 2016) examined website usability, their limited scope (n=3) or national focus failed to address cross-consortia design disparities. Comparative studies, such as Tella and Oladapo's (2016) analysis of electronic resource modules across 20 university libraries in South Africa and Nigeria, or Majid's (2019) cross-national comparison of Indonesian and American university websites, have typically prioritized national contexts over interconsortia perspectives. This gap underscores the need for systematic comparisons of login page designs across library alliances to elucidate functional patterns, tool integration, and application typologies.

To address this, the present study employs a comparative analytical framework to evaluate login page designs across multiple academic library consortia. By identifying best practices and optimizing design principles, this research aims to establish an evidence-based foundation for enhancing user-centric authentication interfaces in academic libraries.Based on these concerns, this study aims to explore whether the login pages deverger from different academic consotia. To guide this investigation, we propose the following research questions:

1. Is there clear pattern or content disparities in library login interface design among consortia?

2. How do functional design choices (e.g., language options, library-specific guidance) align with institutional goals, such as resource access versus cybersecurity?

3. What is the potential reason behind features or authentication methods differ across academic library consortia, and what factors (e.g., alliance maturity, multilingual needs, security priorities) shape these differences?

4. What challenges arise between different functional regime, such as security requirements and usability in library login design, and how do these affect user engagement (e.g., abandonment rates, satisfaction)?

To address these questions, this study investigates the design and functionality of university library login pages, focusing on analysis page content by both quantatitive and literature review appraoches. (Ezell et al.,2022) Our methodology involves a multi-stage approach: Data Collection using screenshots and HTML files, Data Storage, Text Extraction, and Content Analysis employing categorization, statistical analysis, and comparative analysis. The use of screenshots and HTML files in data collection enables researchers to capture both the aesthetic and functional elements of login pages. Furthermore, employing text extraction techniques ensures that critical textual components are collected and analyzed for their relevance to user needs. Content analysis through categorization and statistical methods allows for the identification of patterns in feature implementation across different libraries. This comparative analysis can reveal how various institutions prioritize different aspects in the design. This structured approach allows for a comprehensive understanding of how login page designs can enhance user experience and accessibility (Vasishta, 2013). By employing it, this study expects to contribute valuable insights into the ongoing evolution of library services in the context of continuous progressing cooperation (Barbaresi, 2021).

## Literature Review

In general, university libraries are continuously evolving and corperate in various way to meet the everchanging needs, necessitating a comprehensive forms in whether collaboration or websites design. To some extent, this evolution is driven by technological advancements, societal changings, and the increasing complexity of information access(Balaji, 2019). On the one hand, Associations facilitates resource sharing and expertise pooling, enhancing service delivery and reducing costs for libraries (Bailey-Hainer et al., 2024) On the other hand, it also Decided that the webpage, especially the login design, must meet certain requirements, as they may provide services for identical users.

The literature review is conducted from the progressive relationship of "macro service cooperation mechanism → technology implementation → user behavior → research tools". Starting from the "cooperation model", it gradually transitions to "technology implementation → design principles → analysis methods", analyzing the current research situation, potential research gaps and directions for improvement.

### Collaboration and Partnerships

Academic library consortia have evolved from traditional resource-sharing agreements to multifaceted partnerships encompassing joint acquisitions, vendor collaborations, and cross-border alliances. Early consortia like the Big Ten Academic Alliance (BTAA), established in 1958, pioneered collaborative collection development to reduce redundancy and enhance accessibility (Pionke & Schroeder, 2020). Similarly, Ivy Plus libraries emphasized preserving rare collections while fostering open science initiatives (OCLC, 2022). These mature alliances exemplify institutionalized cooperation driven by shared goals of resource optimization and cost reduction (Bailey-Hainer et al., 2024).

Emerging alliances, such as Sino-foreign Joint Venture Universities (JVUs), represent a paradigm shift. Unlike traditional consortia, JVUs prioritize resource complementarity, integrating domestic and international expertise in curriculum design and technological adoption (Yi, 2020). Their reliance on cloud platforms and multilingual interfaces reflects a hybrid governance model balancing cultural adaptation with global standards (Harrison et al., 2015; Sun & Yang, 2021). However, comparative studies on their login page designs remain scarce, particularly regarding how alliance maturity and regional IT infrastructure shape interface priorities—a gap this study addresses.

**Authentication & Access Management**

Authentication mechanisms are central to balancing security and accessibility in collaborative environments. Federated identity management (FIM) systems, such as SAML-based Single Sign-On (SSO), have become cornerstones for academic libraries, enabling seamless cross-domain access while maintaining privacy (Felts, 2023; Ruenz, 2022). SAML's dominance stems from its ability to unify authentication across diverse platforms, from academic databases to journal repositories (Singley, 2020).

Yet, tensions persist between convenience and security. IP recognition, though widely adopted for on-campus access, limits off-campus usability. Conversely, SSO systems risk complexity for non-technical users, necessitating intuitive interface guidance (Singley, 2020). Emerging alliances like JVUs further complicate this balance by catering to multilingual users and relying on third-party vendors for security solutions—a trend underexplored in existing scholarship (Richardson, 2024).

**Design & Usability**

Effective login page design hinges on minimizing cognitive load while maximizing inclusivity. Cognitive load theory underscores the importance of streamlined interfaces, where features like responsive layouts and clear navigation enhance user engagement (Blessinger & Comeaux, 2020; Raut & Singh, 2024). Universal design principles, such as multilingual support and screen-reader compatibility, are critical for serving diverse populations, including international students and users with disabilities (Redkina, 2024).

However, existing research predominantly examines holistic website usability, neglecting login page specifics (Madhusudhan & Nagabhushanam, 2012). For instance, while Abifarin et al. (2019) analyzed resource accessibility across three libraries, their limited scope precluded cross-consortia insights. Similarly, comparative studies like Tella and Oladapo's (2016) focus on national contexts rather than alliance-level variations. This oversight highlights the need for systematic analyses of how design choices—such as language options or security notices—align with institutional priorities (e.g., JVU's emphasis on IP restrictions vs. BTAA's minimalist interfaces).

## Quantative Analysis in UI Design

Quantitative methods face unique challenges in analyzing login page complexity. Traditional statistical models often assume page independence, a flawed premise given the nested structure of alliance members (Zhu et al., 2007). Hierarchical clustering and functional classification offer partial solutions by grouping features spatially or thematically (Dong et al., 2011). For example, Sharib and Rahman's checklist-based approach provides actionable design benchmarks but struggles to integrate structural and content data. This methodological fragmentation underscores the value of mixed-method frameworks combining visual analysis (e.g., screenshots) with text mining—an approach adopted in this study to capture both aesthetic and functional dimensions (Schmidt et al., 2020).

Combined with previous studies, it is not difficult to find that the logical or visual division of functional types is usually part of the analysis and a necessary step in the comparison. This method connects small functions or elements on the page together based on the similarity of functions or content. This method aims to improve the efficiency and comfort of user operations by optimizing information architecture and interface design. For example, UI elements such as text, controls, and pictures usually cannot exist in isolation, but are grouped to achieve specific interactive functions or visual information, so that the page has consistency and readability, increasing the user's operating comfort (Xiao et al., 2024)

To conclude, recent academic library collaborations have diversified beyond resource-sharing to joint acquisitions and vendor partnerships, exemplified by alliances like BTAA, Ivy Plus, and emerging Sino-foreign Joint Venture Universities (JVUs). These Cross-sector collaborations also extended with commercial entities, such as the ALMA

Consortium, which have profound impact in technology endosement, such as user authentication and login page design. While SSO enhances user convenience, security compromises and interface complexity necessitate user-centric guidance. Design priorities emphasize usability, accessibility, and cognitive load reduction, with responsive layouts, intuitive navigation, and inclusive features (e.g., multilingual support, screen-reader compatibility). Quantitative UI analysis faces methodological challenges due to page complexity; hierarchical clustering and functional classification offer partial solutions but struggle to integrate content and structural data.  Overall, evolving collaborations demand adaptive authentication frameworks, user-centered design, and innovative analytics to optimize resource sharing and accessibility in academic libraries.

## Methodology

In general, this study investigates the consensus and differences in the design of login pages across university alliance libraries, including IVY Plus (13 members), BTAA (17 members), JULAC (8 members), and JVU (15 universities, currently 11 library websites available). The Total sample amount is forty-nine, excluding some universities who restrict public access or affiliate to other institutions. The raw data comes from the official website and is cross-validated with the required data provided by the organization's official website to ensure the maximum coverage and minimum omissions.

According to the relative works, content analysis and comparison are selected as the main method. The methodology is structured into four main phases: data collection, data storage, text extraction, and content analysis, as illustrated in the provided figure. The analysis is based on the results of quantitative comparison combined with qualitative analysis of existing literature and experience to ensure the originality and objectivity of the research.

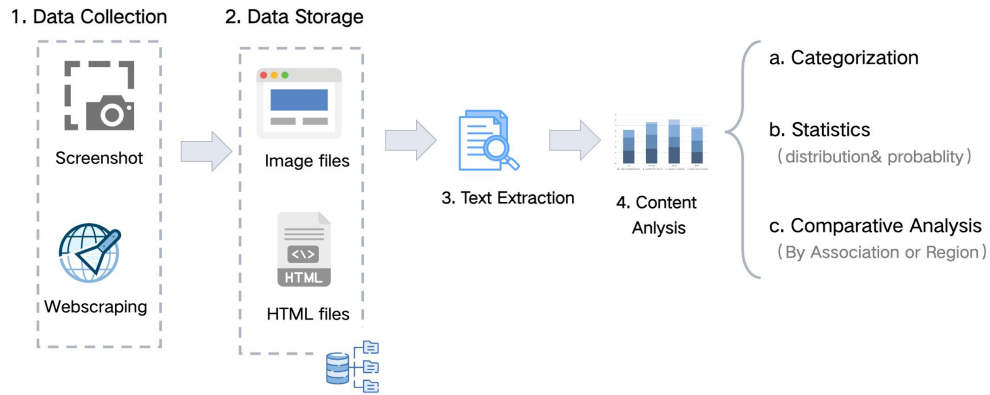

## 1. Data Collection

To comprehensively capture the design features of library login pages, two primary data collection methods are employed:

- Screenshot Capture: Screenshots were captured using FullPage Screen Capture (resolution: 1920×1080) in Chrome Browser, and HTML data was scraped via Selenium (Python 3.10) after ensuring full page loads.
- Web Scraping: HTML content is extracted using web scraping tools to collect structured data about page elements, such as authentication mechanisms and usability features . This approach ensures that both visual and functional aspects of login pages are captured (Schmidt et al., 2020).

## 2. Data Storage

Since the data has been collected by two ways, it is stored in two formats:

- Image Files: Screenshots are saved as image files.
- HTML Files: Scraped HTML data is stored as text files for further analysis. This dual-format storage ensures that both visual and textual elements are preserved for subsequent phases.

## 3. Text Extraction

Textual content from both image and HTML files is extracted using advanced techniques:

- HTML Parsing: HTML files are parsed to identify key elements such as login instructions, authentication fields, and error messages.
- Manual recognition: As the screenshots has been recorded, we can observe the content directly.

This step ensures that all textual components of the login pages are available for detailed analysis.

## 4. Content Analysis

This section will analyze the similarities and characteristics between the same school alliances. Generally, as part of the public presence, university library websites reflect the school's public image. Different schools have different context reveal in design considerations. The extracted data underwent a multi-faceted content analysis process to identify key trends and patterns in university library login page design. This involved a systematic approach incorporating categorization, statistical analysis, and comparative techniques, aligning with established methodologies in library and information science.

a) Categorization: Features of login pages were categorized into four primary groups to facilitate a structured understanding of their functionalities (Mehta，2021). This research method has a certain degree of originality. It further strengthens the statistics of functional categories based on the existing content analysis and comparison. These categories included:

- Authentication Mechanisms: Encompassing methods such as ID/password, email/password, QR code login, and guest login, reflecting the various approaches employed to verify user identity.

- Usability Features: Including elements like "Forgot Password" options, account selection, and language options, which contribute to a user-friendly and accessible interface.

- Security & Compliance: Addressing critical aspects such as privacy policies and copyright information to ensure adherence to legal and ethical standards.

- Library-Specific Features: Incorporating elements like service integration and special instructions tailored to the unique resources and services offered by each library.

b) Statistical Analysis: Quantitative analysis was conducted to determine the distribution and frequency of various features across the sampled libraries. This is previously used in subject services comparison (Hu, 2019), in the context of this artical, it involved employing statistical tools to identify trends in feature adoption, providing insights into the relative prevalence of different design choices.

c) Comparative Analysis: Login pages were compared based on university alliance (e.g., IVY Plus vs. BTAA) and geographic region to identify recurring design patterns and notable differences.

**Findings**

This part aims to provide a brief respone to research quetions and dive into queantative data analysis from various perspectives. Overall, the findings of this article, encompassing statistical analysis of page content and comparative assessments, employ common data analysis and visualization techniques. Initial data organization is followed by statistical methods to analyze and summarize. Overall, university login page designs exhibit both a general consensus and notable differences across alliances. Results indicate that emerging university alliances prioritize website security, often employing access restrictions, while public university alliances, such as BTAA, tend to emphasize discovery and retrieval functionalities for facilating users, de-emphasizing security features on the login page itself. Furthermore, university alliances commonly feature unique functionalities based on their specific nature, such as JULAC's login via assoication. These findings are presented in four sections: Statistics & Categorization, Comparative Analysis & UI Complexity, Distribution of Features, and Consensus & Differences.

**Statistics& Categorization:**

This part aims to provide a basic summary of data collected, and provide basic categorization by textual analysis. Overall, this study cover 46 univerisity from four trending and typical university allience，and there are 21 distinct features observed from our data source (details provided in the appendix), which can be categorized into four types, as shown in Table 1

Table 1: Categorization of Library Login Page Features

| Category | Features | Description |
|---|---|---|
| Authentication Mechanisms | ID/Password, Email/Password, QR Code, Guest Login, Association Login | Methods used to verify user identity. Includes traditional credentials, alternative methods like QR codes, and access for guests or affiliated organizations. |
| Usability & UX Features | Account Selection, Forgot Password, Keep Signed In, Help Options, Alumni Access, Language Selection | Elements designed to enhance the user experience. Includes options for account recovery, persistent login, assistance, role-based access, and language preferences. |
| Security & Compliance | Privacy Policy, Copyright Info, URL Redirection, Security Notices | Features related to data protection, legal compliance, and secure access. Includes links to policies, notices, and secure redirection to prevent phishing. |
| Library-Related Features | Library Login Instructions, QR Authentication, Service Integration | Features specific to library services. Includes guidance on login procedures, integration with library systems, and access to library-specific functions. |

- **Explanation of Categories:**

Authentication Mechanisms: This category encompasses the methods by which the system verifies a user's identity. Traditional username/password combinations are supplemented by options like email/password, which can offer a more user-friendly approach(Palo Alto University, 2023). QR code logins and association logins provide alternative access routes.

- **Usability & UX Features**: These features focus on improving the user experience. Features like "Forgot Password" options are essential for account recovery, while "Keep Signed In" enhances convenience for frequent users(Walla University, 2025). Account selection is important when users have multiple accounts(Walla University, 2025). Help options guide users through the login process (Palo Alto University. (2023)

- **Security & Compliance**: Security is paramount, and this category includes features that protect user data and ensure compliance with legal requirements. Displaying privacy policies informs users about data handling practices. Security notices alert users to potential risks.

- **Library-Related Features**: These elements are specific to the library context. Service integration ensures seamless access to library resources after login (Walla University, 2025).

## Comparrative Analysis & UI Complexity

In general, the university alliances have a common consensus on the functional design of the library login page to ensure basic user needs and security compliance requirements; additionally, they also have their own unique functions, which may be tailored to meet the preferences from different specific users. These differences reflect the different user positioning, service focus and development of each alliance.

| | | JVU | IVY | JULAC | BTAA | | | |
|---|---|---|---|---|---|---|---|---|
| Authentication Mechanisms | | Login with ID and Password | Login with ID and Password | Login with ID and Password | Login with ID and Password | | | |
| | | Login with Email and Password | Login with Email and Password | Login with Email and Password | Login with Email and Password | | | |
| | | Alternative Auth Method (QR, Phone) | Alternative Auth Method (QR, Phone) | Alternative Auth Method (QR, Phone) | Alternative Auth Method (QR, Phone) | | | |
| | | Guest login (organizations, affiliates) | Guest login (organizations, affiliates) | Guest login (organizations, affiliates) | Guest login (organizations, affiliates) | | | |
| | | Keep-signed-in Option | Keep-signed-in Option | Keep-signed-in Option | Keep-signed-in Option | | | |
| | | Alumni-related Option | Alumni-related Option | Alumni-related Option | Alumni-related Option | | | |
| | | | Affliates | Affliates | | | | |
| Usability & UX Features | | Account-selection Option | Account-selection Option | Account-selection Option | Account-selection Option (Use another account) | | | |
| | | "Forgot/Change Password" Option | "Forgot/Change Password" Option | "Forgot Password" Option | "Forgot/Change Password" Option | | | |
| | | URL redirection | URL redirection | URL redirection | URL redirection | | | |
| | | Language Options | | | | | | |
| | | Special Login Instruction (e.g., QR code or other) | | | | | | |
| | | | | Notice/ Alert | Notice about Log-out for Security | | | |
| Security & Compliance | | Help-related Option | Help-related Option | Help-related Option | Help-related Option | | | |
| | | Privacy Policy Information | Privacy Policy Information | Privacy Policy Information | Privacy Policy Information | | | Universal Feature for all Associations |
| | | Copyright Information | Copyright Information | Copyright Information | Copyright Information | | | Distinct Feature |
| Library-Related Features | | | | Association Login | Association Login | | | Common Feature for some of Assoications |
| | | | discovery | | | | | |
| | | | | Library-specific instructions (e.g., Renew books, see checkouts) | | | | |

As, the comparison figure shows that the core features are generally consistent (in grey background), variations exist in specific UX elements and library integrations (highlighted in blue or orange based on its categorization). These findings align with recent academic literature emphasizing the importance of security, usability, and integrated resources in university platforms.

- In terms of Authentication Mechanisms, there seems to be an universal consensus among all universities, such as username, passwords. The figure indicates the use of "Alternative Auth Method (QR, Phone)," which aligns with MFA trends, as Multi-factor authentication (MFA) are increasingly recommended.

- In Usability & UX Features, a certain amount of universities have 'account-option' jump structure (As Figure 1) to allow external or afflicated users to use library services. JVU has its distinct 'language option', as most of members operated in a billigual or even multi-lingual way. 'Speical Login' , such as QR code and mobile authentification, solely appears in JVU, since    is highly accepted in current social media applications, may implicating the tolorance of innovated technology from emerging insititutions. Overall, apart from providing an intuitive navigation to ensure seamless user experience and enhance engagement,    the UX design should be generally user-oritented.

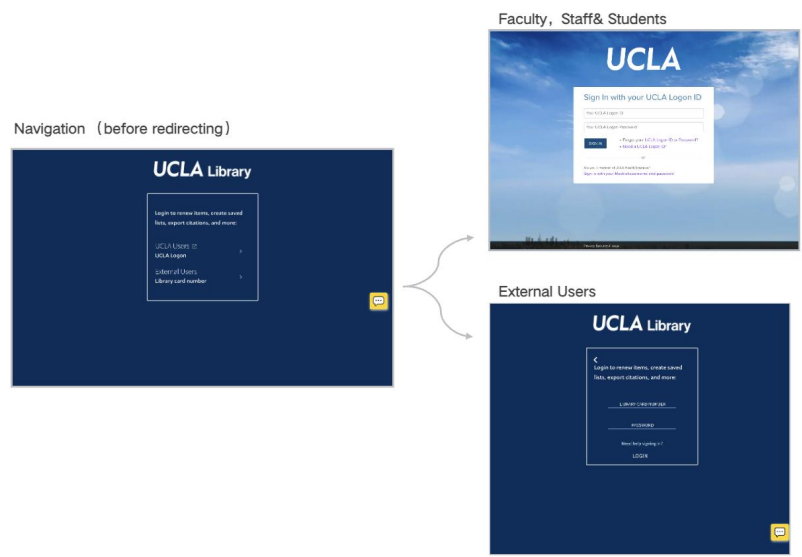

Figure 1. Example of the Jump Login Strucrure

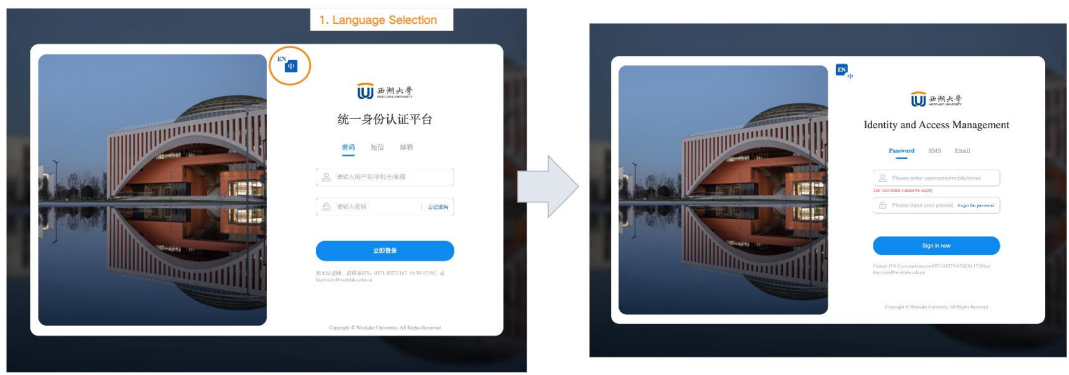

Figure 2. Multi Lingual Option

- Since regulations such as Protecting user data and relevant encryptiuon is essemtial and must comply with privacy regulations (e.g., GDPR, CCPA) (Davis, 2024), There is no obervious variances in Security & Compliance. As the figure lists "Privacy Policy Information" and "Copyright Information," reflecting these prevalent concerns.

- To some extent, Library-Related Features demonstrate the integration of library resources, which is highly correlated to academic support. Particularly in established university alliences (IVY, BTAA and JULAC), Access to digital resources, research databases, and librarian assistance or instructions are conventional and beneficial (Wilson, 2021). The "discovery" feature in IVY and "Library - specific instructions" in BTAA indicate a focus on library integration.

## Distribution of Features Across Consortia

This part examines the general distrubution of features in the design of login pages across university library consortia, including IVY Plus, BTAA, JULAC, and JVU. The findings are derived from two visual analyses: (1) a boxplot showing the distribution of login page features across consortia and (2) a stacked histogram illustrating the average number of features categorized into four functional groups: authentication mechanisms, usability and UX features, security and compliance, and library-related features.

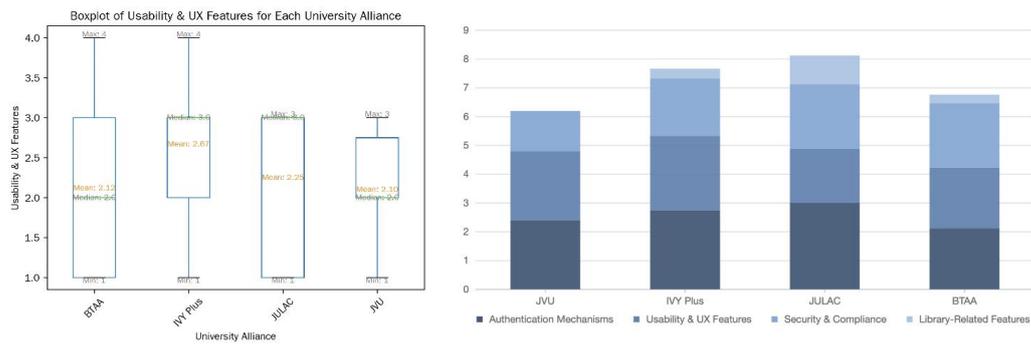

Figure 3. Boxplot and Stacked bar chart of UX Features

The boxplot (Figure 3) reveals variations in the number of login page features implemented by each consortium:

- IVY Plus and JULAC exhibit the highest median number of features (approximately 8), with a broader interquartile range (IQR), indicating significant variability in feature implementation across member institutions.
- BTAA shows a more consistent distribution of features, with a slightly lower median (~7) and a narrower IQR. This suggests a more standardized approach to login page design among BTAA libraries.
- JVU has the lowest median (~6) and the smallest IQR, reflecting limited feature diversity and a more uniform but minimalistic design approach.

These suggest that IVY Plus and JULAC prioritize feature-rich designs, potentially catering to diverse user needs, while JVU adopts a simpler interface.

Regarding the histogram, different alliances have their own characteristics in the distribution and emphasis of specific functions. From the histogram of function distribution, we can see that the specific functions (expressed in code) that frequently appear in each alliance are different. This phenomenon may imply different design concerns. Each alliance may design the login page based on its own user characteristics, service positioning, resources and other factors. For example, some alliance user groups may be highly dependent on library resources, so they pay more attention to the setting of library related functions; Some alliances may pay more attention to the convenience and security of user login, so they invest more in authentication mechanism, usability and user experience functions.

**General consensus & Differences (Functions shared by all alliances)**

This section interprets a heatmap illustrating the average proportion of various type of features on login pages across different university alliances, highlighting notable variations even in commonly implemented features like authentication mechanisms and privacy compliance.

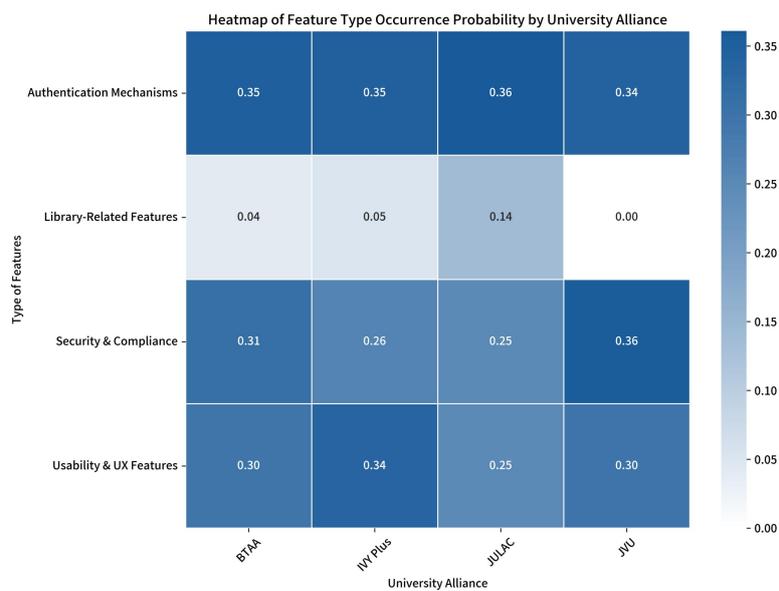

Figure 4. Percentage Heatmap of Features

The heatmap chart summarizes the frequency of functions in different university alliances, and the thermodynamic chart shows the probability of different functions in different university alliances. The color changes from white to blue. The darker the color, the higher the probability of the function in the corresponding league. Through this figure, we can visually see the differences in function settings of each alliance.

From the perspective of consensus, the heat map shows that the probability of each alliance appearing in the authentication mechanism, security compliance, usability and user experience functions is relatively high, indicating that these function types are the basic parts that all alliances generally attach importance to, reflecting the common needs of ensuring user login security and providing good use experience.

Specifically, the BTAA exhibits the lowest proportion of authentication features at 31%, while JVU shows the highest at 36%. Given the relatively stable number of authentication-related features across alliances (averaging between 2.5 and 3.0), this proportional difference stems from variations in the quantity of other features, such as Usability & UX Features and library-related features, aligning with prior analytical findings.

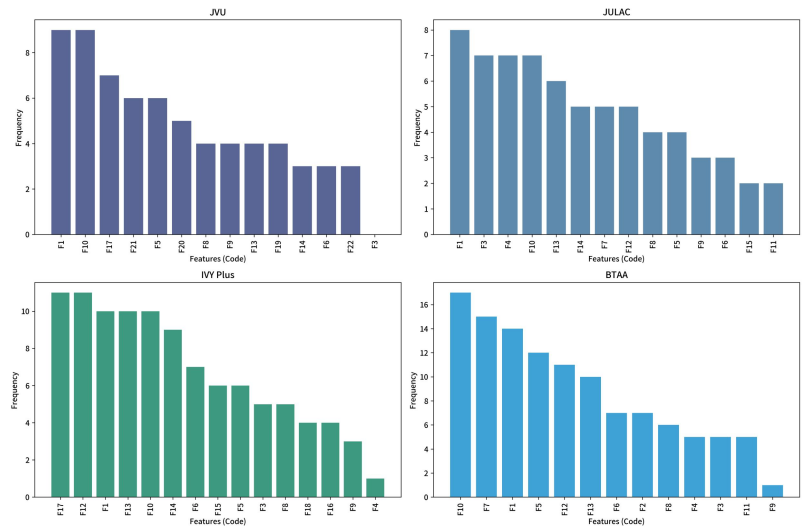

Figure 5. Feature Distribution of University Alliences ( see apendix for feature details)

According to the analysis of variance, these differences may be primarily driven by Usability features, especially the language options (F20) and multifaceted login methods, which is prevalent in one association but is rare in other, as shown in the subsequent figure, suggesting a correlation with multilingual academic environments and varying IT technology landscapes across regions.

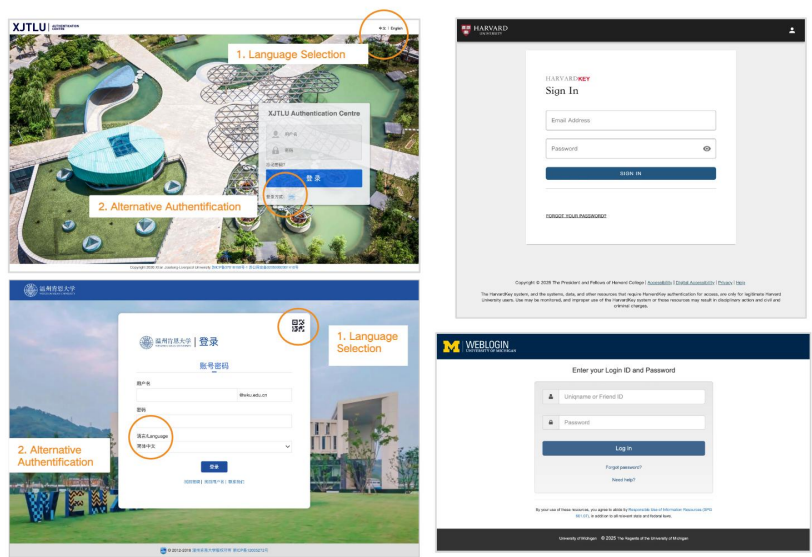

Figure 6. Language Selection& Alternative Authentificaiton

Discrepancies in compliance content are even more pronounced, with JVU exhibiting the highest proportion at 36% and JULAC the lowest at just 31%. A comparative analysis of two specific login pages, the Shenzhen MSU - BIT University and The City University of Hong Kong, reveals distinct differences. The former includes IP restrictions, copyright attribution, registration details, and vendor information; the latter provides university copyright information and access demographics only. Restricting access is a common method for enhancing system security, which might explain why newer universities favor IP restrictions (Nagra, 2019). Additionally, due to limited technical support for in-house development, relying on established vendor solutions is a common practice for university libraries, particularly among emerging institutions.

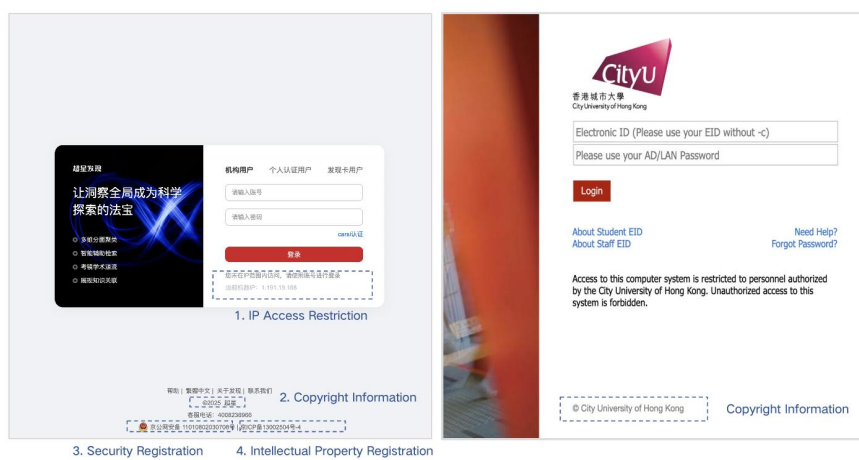

Figure 7. Login Restrictions

These findings potentially correlate with the alliances' establishment dates (BTAA in 1958, IVY Plus in 2006, JULAC in 1967, and JVU in 2009). In general, mature university alliances are more inclined to share resources, enabling broader access to their collections and increasing circulation, resulting in relatively streamlined login pages. In contrast, emerging university libraries tend to prioritize the cybersecurity of their services. Emerging universities usually restricting access as relative public accessability tend to bring more risks. Although it can help improve security, this method also leads to a compromise in user experience (Moorthy&Dietsch, 2023). In addition, too strict restrictions may hinder cirrculation efficiency, affecting academic resource usage (Wilson et al., 2019) Therefore, when designing an access control system, a balance should be found to ensure security while minimizing inconvenience to users (Yang, 2021)

**Response To the Research Questions**

This section presents the study's findings, structured according to the research questions to ensure clarity in how the collected data supports or challenges the initial hypotheses.

**RQ1: Design Patterns and Disparities**

According to the Average Feature number Chart, Consortia exhibit distinct design priorities: JVU emphasizes multilingual support (100% F20) and alternative authentication (45% F19), while IVY and JULAC integrating more library-specific. BTAA's universities have more balanced design among different function regimes, showing that it serves more diverse user groups. These disparities, to some extent, shows institutional goals and user demographics.

Table 1. Average Number of Features

|  | Authentication Mechanisms | Usability & UX Features | Security & Compliance | Library-Related Features |
|---|---|---|---|---|
| JVU | 2.40 | 2.10 | 3.50 | 0.00 |
| IVY PLUS | 2.75 | 2.67 | 2.08 | 0.42 |
| JULAC | 3.25 | 2.25 | 2.25 | 1.25 |
| BTAA | 2.53 | 2.12 | 2.24 | 0.19 |

**RQ2: Functional Alignment with Goals**

There are two main divergences in the library login page design, sercurity and library-related functions. For JVU libraries IP restrictions is prelvant ( 100% ), in contrast to other organizations, which try to flowrish and simplify logins by adding partnerships and affliates, reflecting cybersecurity versus. accessibility priorities. For the library-related services, which barely exist in BTAA and JVU, are more popular in IVY PLUS and JULAC, indicating that research oreiented organizations tend to add discover or instructions in login pages.

**RQ3: Drivers of Feature Differences**

Overall， alliance maturity and user demographics shape designs. Mature consortia (BTAA, IVY PLUS, JULAC) prioritize usability ( having higher average numbers of Usability & UX Features), while newer ones (JVU) focus on security. Also, multilingual needs (JVU's F20) and regional regulations also influence choices. However, JVU's IP restrictions enhance security but risk reducing accessibility, while BTAA's minimalism improves usability but underemphasizes security. Quantitatively, JVU's 2.5 fewer features suggest simplified workflows, but third-party vendor reliance introduces complexity.

## Discussion

### 1. Common Design Consensus for University Library Login Pages

The study reveals that the design of university library login pages shares certain functional commonalities across institutions. All alliances prioritize core authentication mechanisms (e.g., username/password, email/password), security and compliance (e.g., privacy policies, copyright information), and user experience features (e.g., "Forgot Password" options, "Stay Logged In" functionality)（Zulfiqar, A., & Khalid, A. (2024).）. These features are considered "common knowledge" in login page design, reflecting the fundamental requirements for ensuring user login security and providing a positive user experience.

Moreover, there are notable differences in functional settings among alliances. For example, some alliances emphasize library-related functionalities (e.g., guidance on resource renewals), while others prioritize convenience and security during the login process, such as JVU put an emphaize on multi-national corperation tend to have 'multi-language' and 'alternative login' options, which indicates that the relative open attitude toward emerging technologies of those new institutions. At the same time, some new established institutions tend to scarify accessability for security by visit controlling, such as IP restriction or simplified login pages, rather than accessibility for security concerns. These differences reflect variations in user group targeting, service priorities, and developmental focus among alliances in different stage of development.

### 2. Current Trends in Library Login Page Design

To some extent, the analysis reveals distinct trends shaped by institutional priorities, technological adoption, and organizational maturity. First, security-centric design practices dominate emerging university alliances (e.g., JVU, founded in 2009), where IP restrictions and third-party vendor solutions are prevalent. This aligns with modern cybersecurity demands, particularly for institutions lacking in-house technical expertise. Conversely, older alliances like BTAA (established in 1958) prioritize resource accessibility or cirrculation over security, offering streamlined login pages to reduce authentication barriers and leave more space for dicovery or library-related functionalities. This phnoemenen in well-developed universities reflects the historical and ultimate mission for university libraries is to maximize resource circulation across member institutions.

A second trend is the divergence in feature distribution driven by regional and functional needs. For instance, JVC's universal login system caters to its multilingual academic environment, while variances in privacy compliance (e.g., JVU's 39% vs. JULAC's 23%) highlight differing legal or institutional priorities and JULAC libraries generally have alliance member login functions. Different technoecological genomes influence how systems are designed and used (Ex Libris., 2024).

## 3 Limitations of Existing Practices

While this study identifies critical design tensions in contemporary library services, its findings underscore the necessity of transcending disciplinary and methodological silos. By bridging qualitative insights with quantitative validation and leveraging cross-disciplinary theories, subsequent research can advance toward more holistic, evidence-based interface optimization strategies. The analysis reveals persistent tensions in optimizing library service interfaces, particularly in reconciling functional complexity with user-centric simplification. Two critical dichotomies emerge:

### 3.1 Security-Centric Complexity Versus Usability Trade-offs

The research data reveals the explicit tension between security protocols and user experience. Taking JVU as an example, although its high proportion of security features (such as IP restrictions) reduces the risk of unauthorized access, it also leads to an increase in interface complexity (the median number of features is 6, IQR 1.2), which may increase users' cognitive load and task abandonment rate. On the other hand, although BTAA's minimalist design improves usability, its security feature ratio (23% F12) is significantly lower than that of other alliances, implying data leakage risks. This contradiction is further explained under the framework of cognitive load theory: interface complexity is negatively correlated with user task efficiency (Blessinger & Comeaux, 2020), but the lack of security features may weaken user trust.

To resolve this contradiction, this study proposes a "dynamic security adaptation" framework: dynamically adjust security strictness through context-aware technology (such as multi-factor authentication based on geographic location). For example, on-campus users can enable low-friction SSO authentication, while off-campus access triggers biometric verification. This type of design has been proven to be effective in the field of e-commerce, and its impact can be quantified through analysing massive user behavior records in library scenarios after the library service operate in a stable and merely changing manner.

**3.2 Analytical Scope and Behavioral Granularity**

While the study quantified interface features (e.g., security element frequency), it lacked finer-grained insights into user decision-making processes. Metrics such as task abandonment rates, time-on-task efficiency, or post-interaction satisfaction indices could elucidate whether observed design patterns align with user expectations. Incorporating behavioral analytics tools, such as A/B testing or heuristic usability evaluations, would enable a more nuanced assessment of design efficacy.

# 4 Future Directions for Research and Design

**4.1. Adaptive Security-Usability Frameworks**

Although this paper lacks an analysis of specific authentication technology selection, it is basically certain that most university libraries differentiate between on-campus users, i.e., staff students, and other users, and that on-campus users favor CAS authentication login methods, which is conventional but can not deny future possibilities. Future research could prioritize the development of adaptive authentication technolyies like biometric authentication (e.g., fingerprint or facial recognition) and passwordless login methods (e.g., WebAuthn) offer opportunities to enhance security without sacrificing usability. In addition, libraries may also explore AI-driven anomaly detection systems to identify and mitigate potential security threats while minimizing false positives that disrupt user workflows.

**4.2. Cross-Disciplinary User Behavioral Analytics**

It's worth noting that the current way of analyzing and designing pages is on the conservative side, to address the limitations of current analytical approaches, future studies could integrate behavioral data collection methods, such as A/B testing, heuristic evaluations, and eye-tracking, to measure user engagement and task efficiency. For example, tracking abandonment rates during multi-step authentication processes could reveal pain points in interface design. Additionally, mixed-methods research combining quantitative feature analysis with interviews would provide deeper insights into how design choices align with user expectations, particularly among diverse populations like international students or alumni.

### 4.3. Standardized yet Flexible Design Templates

Functional differences across alliances reveal divergences in resource allocation and service positioning. Mature alliances meet diverse needs through functional redundancy (IVY median 8 features), while emerging alliances rely on standardized templates for rapid deployment. This suggests that open source modular design templates may be the key to balancing consistency and flexibility. For example, alliances can jointly build a basic template that includes core authentication and multi-language support, allowing members to add security or resource integration modules as needed. Such collaboration requires innovation in technical governance mechanisms between alliances, such as establishing a cross-institutional UI design working group and formulating interoperability standards.

Given the disparities in feature implementation across consortia, collaborative efforts to develop open-source design templates could streamline interface development while accommodating regional and institutional variations. These templates could include modular components allowing libraries to customize interfaces without reinventing foundational elements. Further, it is allowed to make adjustments to the page functionality by quantitatively analyzing user behavior to make the page content more responsive to the user's needs, or the page content can be customized by an authenticated user.

## Conclusion

This study evaluates design paradigms of university library login pages across prominent academic alliances, revealing both shared principles and divergent priorities shaped by institutional objectives, technological adoption, and regional contexts. The analysis of 46 institutions within IVY Plus, BTAA, JULAC, and JVU underscores a foundational consensus in core functionalities — authentication mechanisms, security compliance, and usability features — reflecting universal imperatives to balance accessibility, security, and user experience. However, significant variations emerge in feature emphasis, with mature alliances like BTAA prioritizing resource discovery through streamlined interfaces, while emerging consortia such as JVU emphasize cybersecurity measures like IP restrictions and third-party vendor integrations. These distinctions highlight the tension between historical missions (e.g., maximizing resource circulation) and contemporary demands for digital safeguarding.

Notably, regional and linguistic diversity profoundly influences design choices. JULAC's universal login system and multilingual support cater to its cross-border academic environment, whereas variances in privacy policy prominence signal differing institutional or legal priorities. Statistical analyses further identify usability features—particularly multilingual options and multi-method logins—as the most significant drivers of cross-alliance disparities, suggesting that user-centric adaptations are increasingly tailored to localized ecosystems. Such findings align with broader trends in library science, where federated identity management and responsive design are critical to serving heterogeneous user bases.

The study also exposes critical limitations, including functional overload in security-centric designs and underprioritized authentication in legacy systems. These challenges underscore the need for adaptive frameworks that harmonize security with simplicity. Future research should integrate behavioral data and mixed-method approaches to evaluate how design choices impact user engagement and task efficiency. Additionally, cross-alliance collaboration could foster standardized yet flexible templates, leveraging open-source solutions to reduce vendor dependency. Innovations such as context-aware multi-factor authentication and AI-driven personalization may further enhance inclusivity without compromising security.

Future research may focus on integrating mixed technological innovations to overcome current limitations. First, A/B testing platforms (e.g., Optimizely) can be used to compare interfaces (e.g., on-campus SSO vs. ip restrictions), measuring task completion rates to quantify user experience. Second, machine learning models can analyze user behavior logs (e.g., Apache Kafka stream data) to improve operational mechanisms—such as using gradient-boosted decision trees to automatically hide non-essential features and reduce cognitive load. Additionally, an open-source collaborative ecosystem is recommended, such as building a modular UI template library on GitHub that includes GDPR-compliant privacy statements and multilingual selectors. This would allow consortia to flexibly customize interfaces within a standardized framework, balancing security, usability, and regional needs. These directions may enhance the intelligence of current portals,  and user experience.

# Appendix

Table 1. An index table of all features

| Index | Feature |
|---|---|
| F1 | "Forgot/Change Password" Option |
| F2 | Account - selection Option (Use another account) |
| F3 | Alumni - related Option |
| F4 | Association Login |
| F5 | Copyright Information |
| F6 | Guest login (external organizations) |
| F7 | Help - related Option |
| F8 | Keep - signed - in Option |
| F9 | Login with Email and Password |
| F10 | Login with ID and Password |
| F11 | Notice/ Alert |
| F12 | Privacy Policy Information |
| F13 | URL redirection |
| F14 | Account - selection Option |
| F15 | Affliates |
| F16 | Alert |
| F17 | Help - related Option（contact） |
| F18 | Library - specific instructions (e.g., Renew books, see checkouts) |
| F19 | Alternative Auth Method (QR， Phone) |
| F20 | Language Options |
| F21 | Legistration Information (Privacy Policy etc) |
| F22 | Special Login Instruction (e.g., QR code or other) |